\documentclass[11pt]{article}
\usepackage{moriond,epsfig}

\def\Journal#1#2#3#4{{#1} {\bf #2}, #3 (#4)}

\def\NPB{{\em Nucl. Phys.} B}
\def\PLB{{\em Phys. Lett.}  B}

\def\PRD{{\em Phys. Rev.} D}

\def\be{\begin{equation}}
\def\ee{\end{equation}}
\def\bea{\begin{eqnarray}}
\def\eea{\end{eqnarray}}

\begin{document}
\vspace*{4cm}
\title{HEAVY FLAVOUR PRODUCTION AT HERA}

\author{ S. MIGLIORANZI \\
for the H1 and ZEUS collaborations }

\address{DESY, 85 Notkestrasse,\\
Hamburg 22607, Germany}

\maketitle\abstracts{
Studies of charm and beauty production in {\it ep} collisions with a center-of-mass 
energy of 318 GeV are reported from the two HERA collaborations, H1 and ZEUS. The analyses 
make use of both the HERA-I data sample recorded between 1996 and 2000 and a sample from HERA-II, which started in 2003. 
The cross sections measured by both H1 and ZEUS experiments are compared 
with next-to-leading order QCD calculations. The measurement of the charm and beauty contributions to the 
proton structure function is also presented. The comparison to next-to-next-to-leading order (NNLO) calculations shows agreement within the errors.
}

\section{Introduction}

Heavy quark production processes provide a powerful insight into the understanding of Quantum Chromodynamics. The large mass of 
the heavy quark makes the perturbative calculations reliable, even for total cross sections, by cutting off infrared singularities 
and by setting a large scale at which the strong coupling can be evaluated. At HERA heavy flavour production is possible both in 
photoproduction and deep inelastic scattering (DIS) reactions, the latter having dramatically smaller cross sections. While 
in photoproduction the photon virtuality $Q^2$ is very small ($Q^2 \sim 0$), and the photon is almost real, in DIS $Q^2$ can reach values much higher 
than the squared quark mass $m^2_q$. In direct-photoproduction processes the quasi-real photon enters directly in the hard interaction whilst in resolved-photoproduction processes the photon acts as a source of partons that take part in the hard interaction.\\
%The first measurements of beauty production at HERA\cite{oldzeus,oldh1} revealed a slight excess of the data over pQCD predictions. Si
%milar findings were made in $p \bar{p}$ and $\gamma \gamma$ collisions at Tevatron\cite{tevatron} and at LEP\cite{lep}. The most recen
%t HERA results\cite{muj_h1,muj_zeus,life_h1,dstar_h1,longhin,dimu_ZEUS,my_ZEUS} are reported here.
%\section{Heavy Quark Production}
The dominant process for heavy quark production in DIS and in direct photoproduction in $\it{ep}$-collisions at HERA is the boson-gluon fusion 
(BGF) mechanism, $\gamma g \rightarrow Q \bar{Q}$. 
%\begin{figure}[htb]
%\includegraphics[width=0.35\textwidth]{./diagr_feyn.eps}
%\caption{Examples of beauty production processes in leading order pQCD.}
%\label{bgf}
%\end{figure}
At leading order (LO), the BGF process is directly sensitive to the gluon content of the proton. In resolved photoproduction it is necessary to also consider quark excitation diagrams, $Q g \rightarrow Q g$, where the heavy quarks originate from the photon, and the gluon-gluon fusion, $g g \rightarrow Q \bar{Q}$. 
In photoproduction both the direct and the resolved components contribute to heavy quark production.\\
Calculation tools are available up to next-to-leading order ($\alpha^2_s,NLO$) in the form of Monte Carlo integration programs \cite{mcp1,mcp2}. They use the massive scheme \cite{massive} in which $u$, $d$ and $s$ are the only active flavours in the proton, and charm and beauty are dynamically produced in the hard scatter. In another (massless) approach, beauty and charm are treated as massless and included in the PDFs. The HERA measurements shown here are compared to massive and massless NLO QCD predictions. 
%Other approaches use the GMVFNS or the ZMVFNS, in which the number of active flavours in the proton varies...

\section{Charm Production}

\begin{figure}[h!]
\begin{center}
\includegraphics[width=0.28\textwidth]{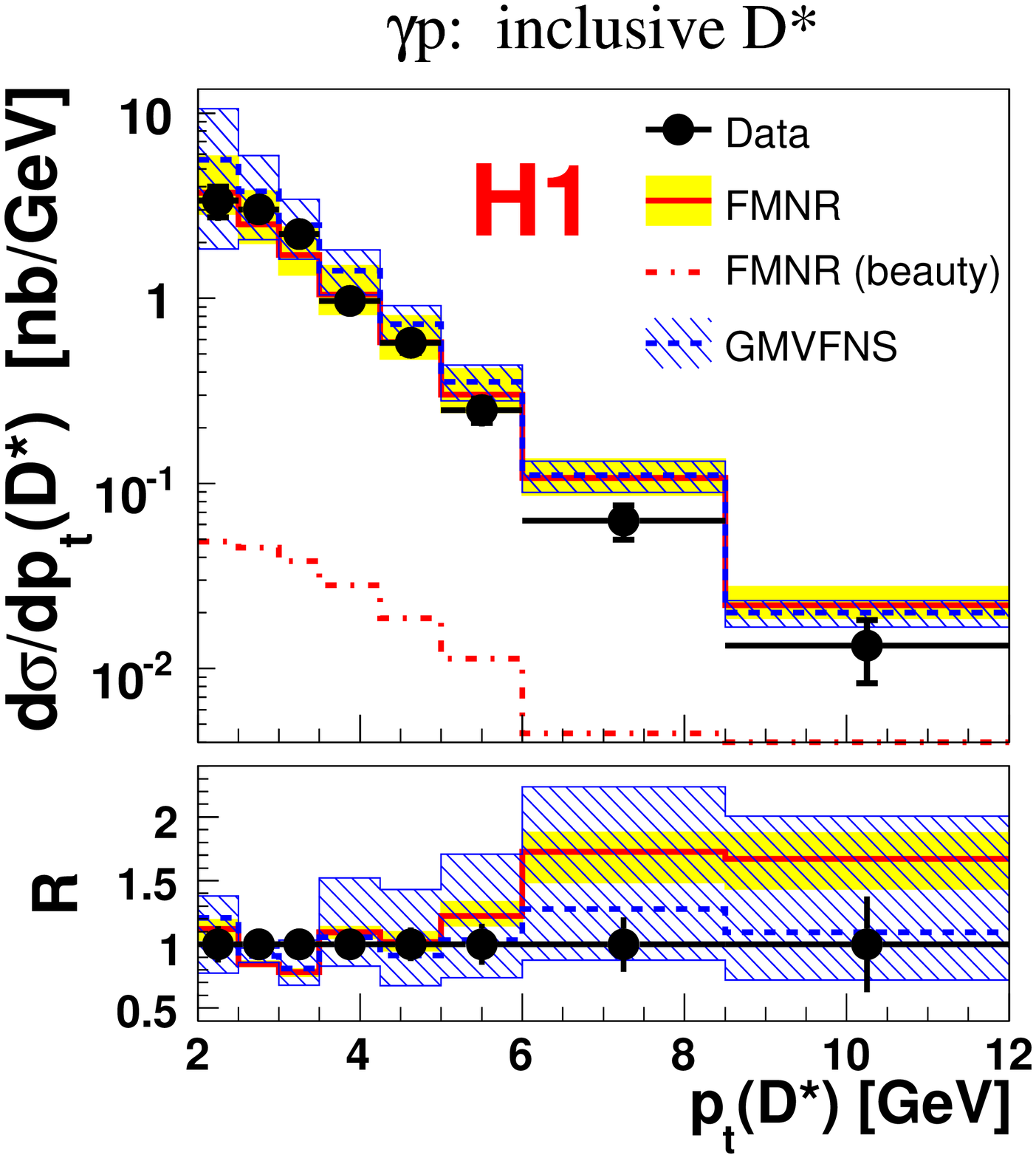}
\includegraphics[width=0.28\textwidth]{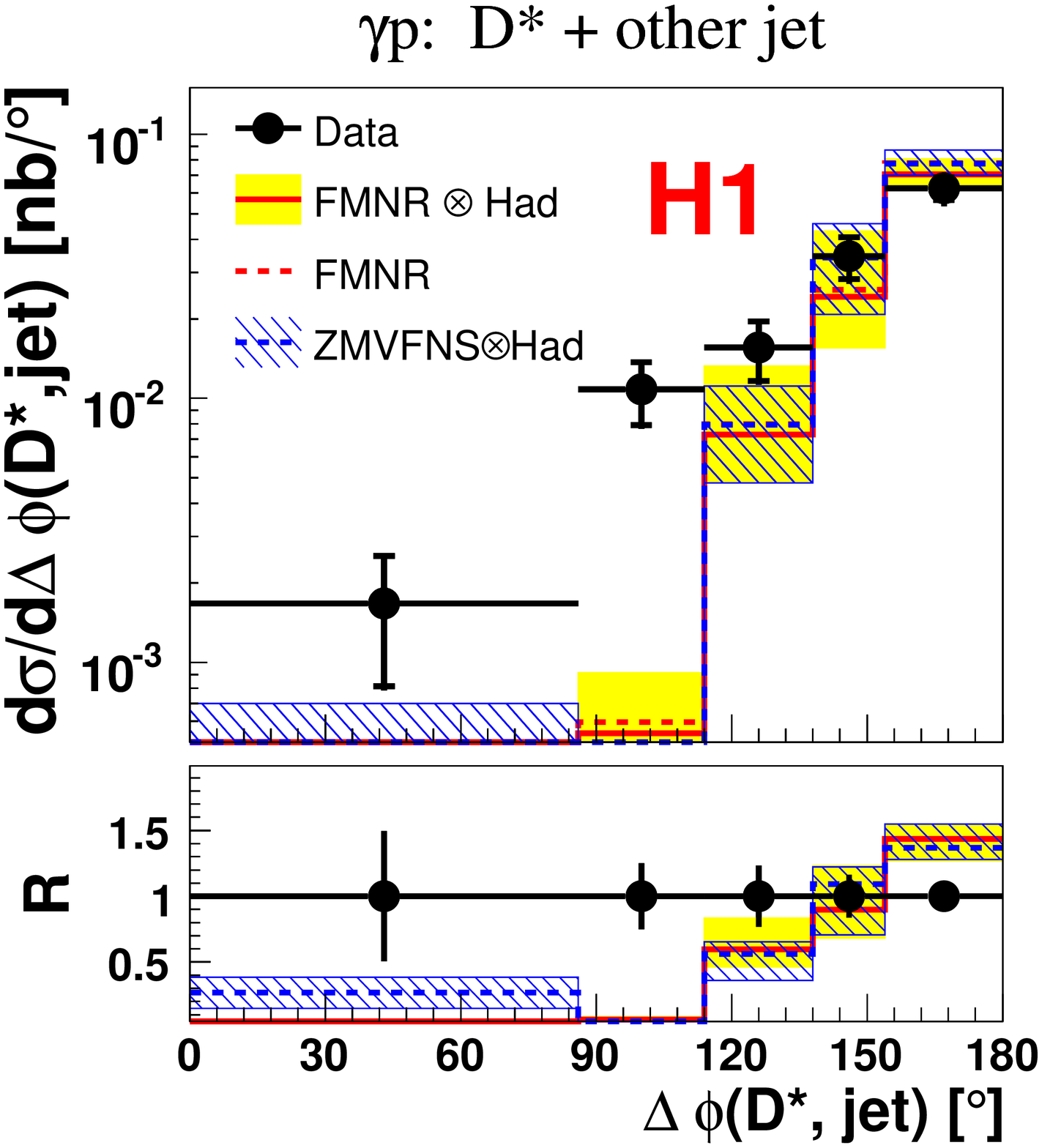}
\caption{(left) Inclusive $D^*$ cross sections as a function of $p_T(D^*)$ compared with the NLO predictions based on FMNR and GMVFNS. (right) $D^*+jet$ cross section as a function of $\Delta \phi(D^*,jet)$ compared with FMNR and ZMVFNS.}
\label{charm_php}
\end{center}
\end{figure}

\begin{figure}[htb]
\begin{center}
\includegraphics[width=0.35\textwidth]{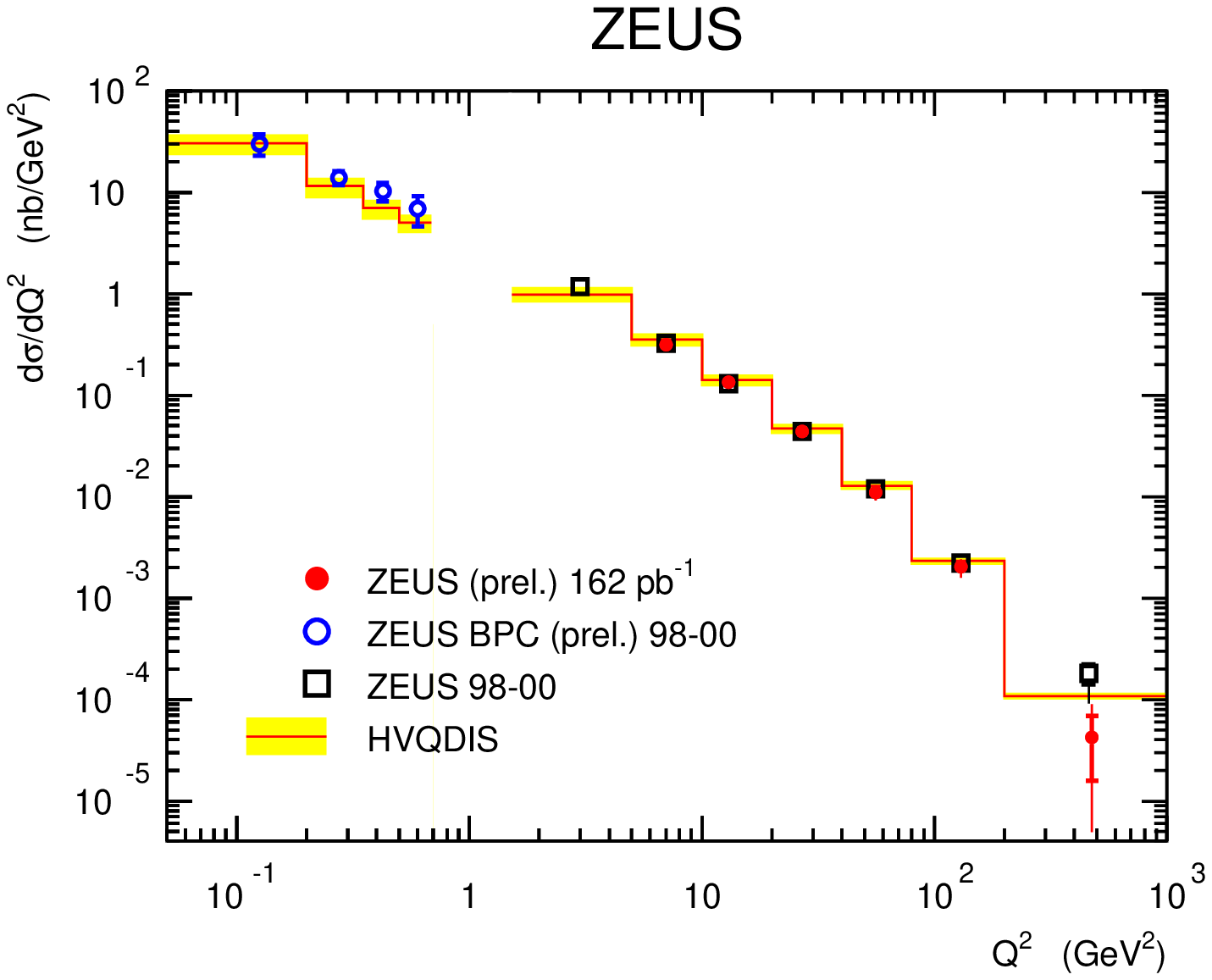}
\includegraphics[width=0.45\textwidth]{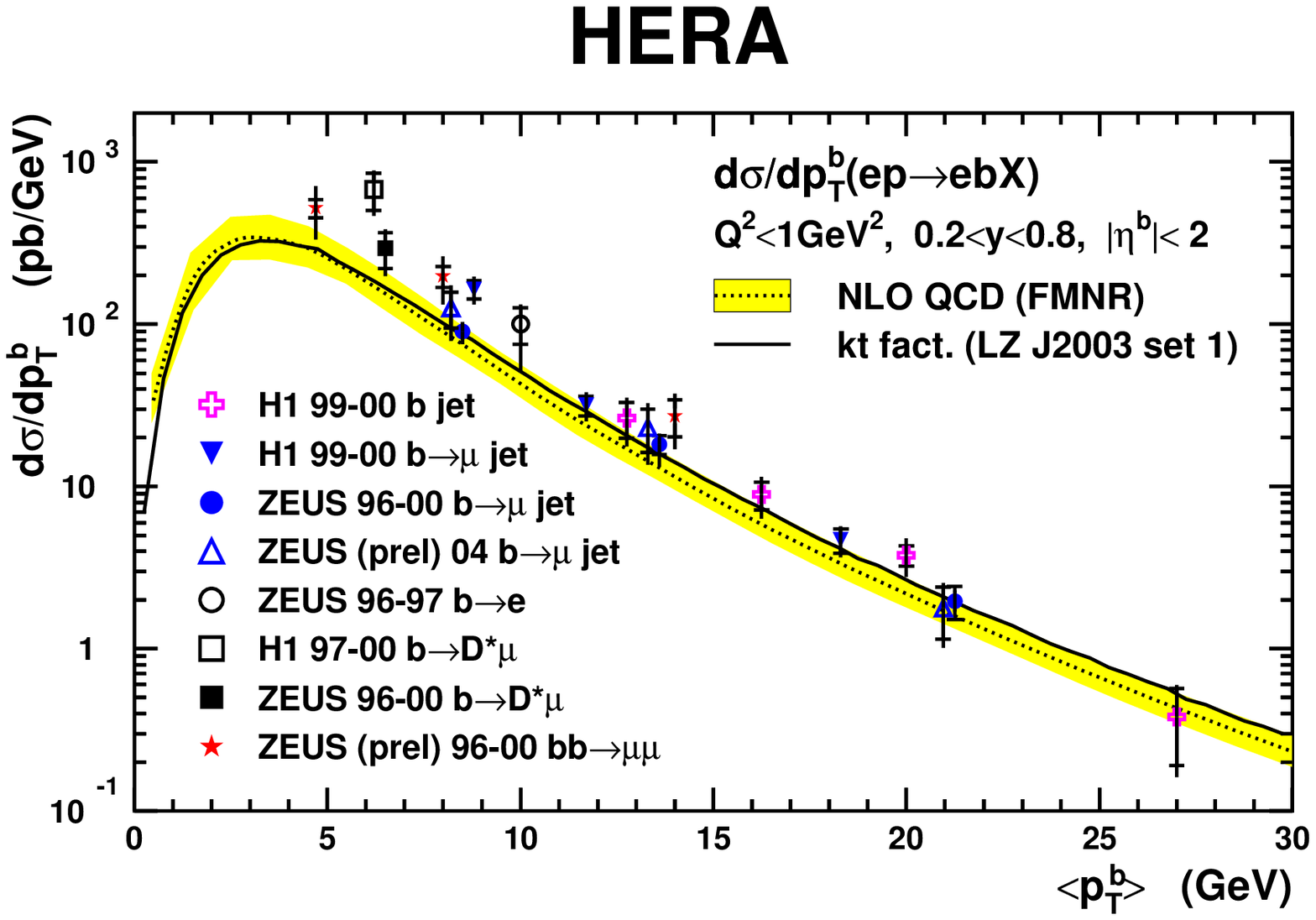}
\caption{(left) Differential $D^*$ cross section as a function of $Q^2$ compared to the NLO calculation of HVQDIS. The HERA-II data (solid points) are shown compared to the most recently published ZEUS measurements (open squares). 
%The inner error bars indicate the statistical uncertainties and the outer bars show the statistical and systematic uncertainties added in quadrature. 
The solid line gives the predictions from the ZEUS NLO QCD fit for $m_c=1.35$ ${\rm GeV}$ with the shaded band indicating the uncertainty in the prediction. The results using the BPC data are also reported (open circles). On the right the summary of the latest beauty cross section measurements using different tagging techniques by ZEUS and H1 
is shown.}
\label{charm_dis}
\end{center}
\end{figure}

The most recent results by the H1 collaboration regarding the $D^{*\pm}$ photoproduction\cite{c_php_pres} used a data sample five times larger than in the previous pubblications \cite{c_php}. The $D^*$ meson was detected via the decay channel~\footnote{Charge conjugate states are implicitly implied.} $D^{*+} \rightarrow D^0 \pi^+_s \rightarrow K^- \pi^+ \pi^+_s $.  Details of the heavy quark production process were investigated by studying events with a jet not containing the $D^*$ meson ($D^* + jet $). In Fig. \ref{charm_php}(left) the measured differential cross section as a function of the transverse momentum of the $D^*$, $p_T(D^*)$, is compared to NLO calculations based on the massive scheme \cite{fmnr1,fmnr2} (FMNR) and a general-mass variable-flavour-number scheme \cite{gmvfns1,gmvfns2} (GMVFNS). The cross section falls steeply with increasing $p_T(D^*)$ as predicted by all calculations. 
%Measurements of correlations between the $D^*$ and the jet have been also performed which are sensitive to higher ordere effects and to the longitudinal and transverse momenta of the partons entering the hard scattering process. 
In Fig. \ref{charm_php}(right) the differential cross section as a function of the difference in the azimuthal angle between the $D^*$ and the other jet $\Delta \phi(D^*,jet)$ is shown. The results are compared with NLO prediction from FMNR and with predictions based on the zero-mass variable-flavour-number scheme \cite{zmvfns1,zmvfns2} (ZMVFNS). A large fraction of the produced $D^*+jet$ combinations deviates from a back-to-back configuration. The NLO calculations are not able to describe the small $\Delta \phi$ behaviour of the data, indicating the presence of higher order contributions in this particular region.\\ 
\noindent
Recent results on $D^*$ production in DIS by the ZEUS Collaboration\cite{c_dis_pres} are
shown in Fig. \ref{charm_dis}(left). The differential cross section in $Q^2$ is compared to the previous HERA-I result\cite{c_dis_old}. Good agreement is seen for the entire $Q^2$ range over which the differential cross section falls by about four orders of magnitude. The cross section is reasonably well described by the NLO calculation which use the ZEUS NLO QCD fit.
In (Fig. \ref{charm_dis})(left) also shown are the ZEUS results on the $D^*$ cross section in the range $0.05<Q^2<0.7$ ${\rm GeV}^2$ \cite{bpc}. The beampipe calorimeter of ZEUS was used for the measurement of the scattered lepton, which allows the first measurement of the transition region between photoproduction and DIS. The NLO calculations describe well also this region.

\section{Beauty Measurements}

\begin{figure}[h]
\begin{center}
\includegraphics[width=0.35\textwidth]{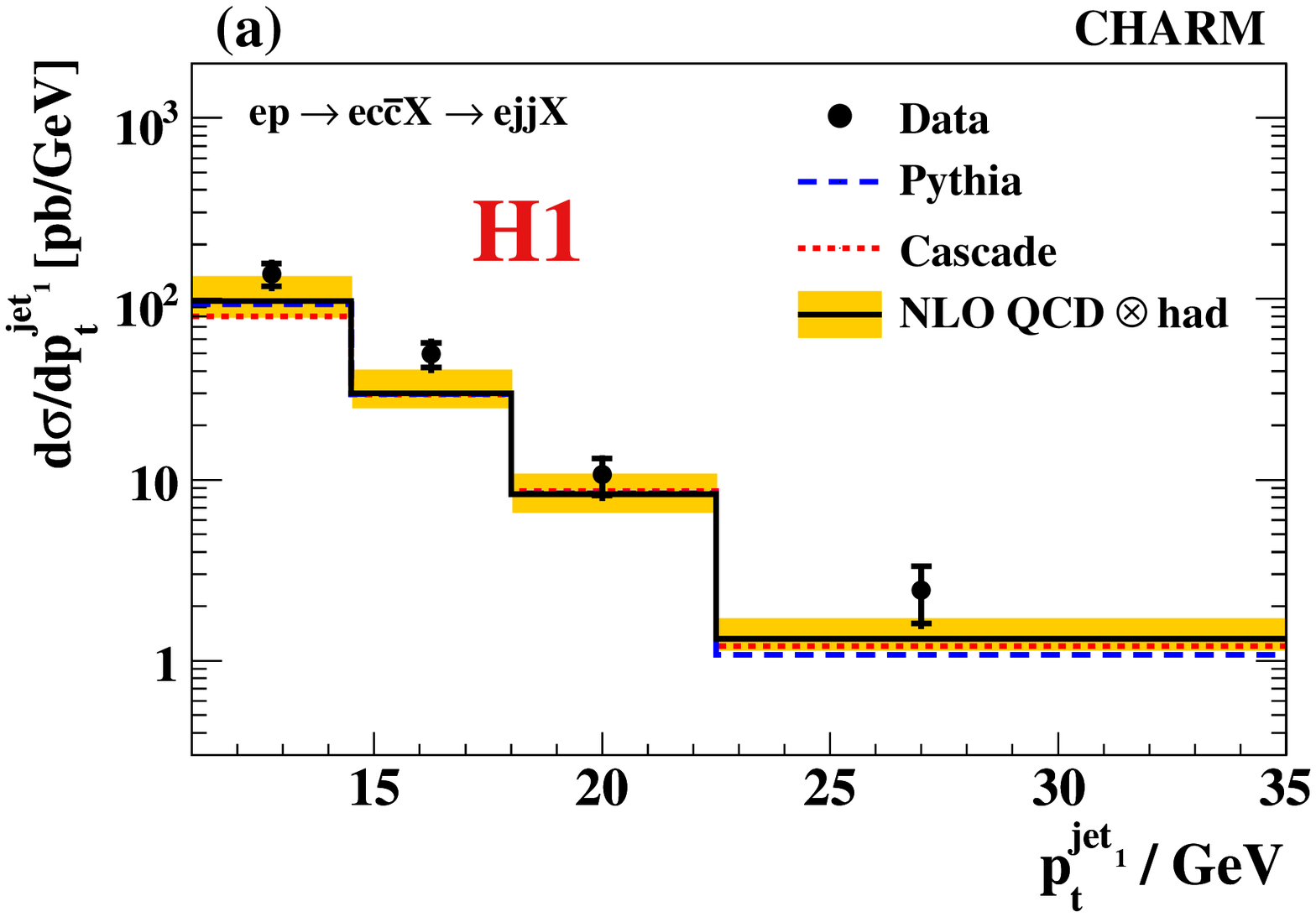}
\includegraphics[width=0.35\textwidth]{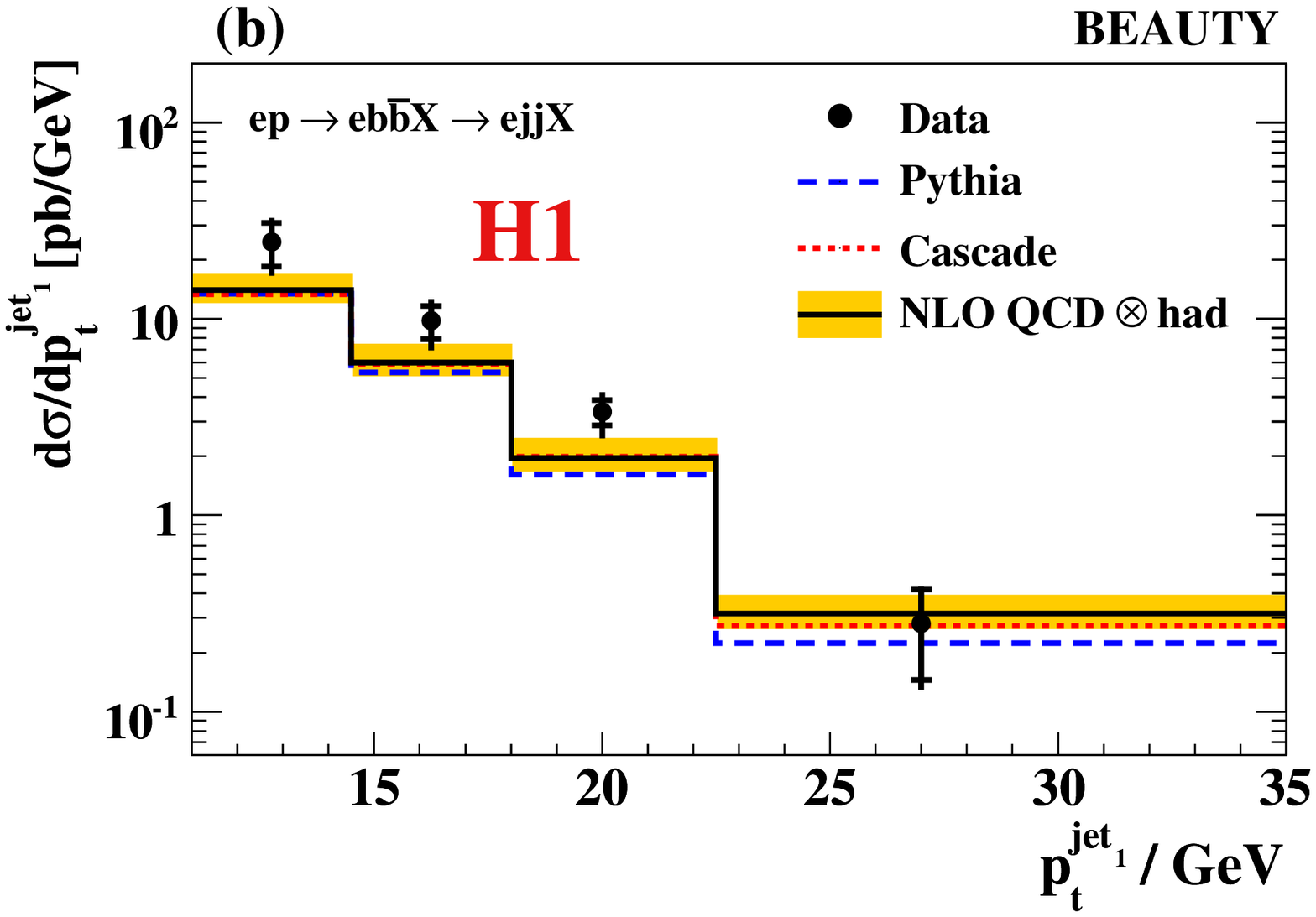}
\caption{Differential charm and beauty PHP cross sections $d \sigma /dp_t^{jet1}$ for the process $ep \rightarrow e(c \bar{c} \; or \; b \bar{b}) X \rightarrow ejjX $. 
%The inner error bars indicate the stastistical uncertainty and the outer error bars show the statistical and systematic errors added in quadrature. 
The solid lines indicate the prediction from NLO QCD, corrected for hadronization effects, and the shaded band shows the estimated uncertainty. The abolute predictions from PYTHIA and CASCADE are also shown.} 
\label{lifet}
\end{center}
\end{figure}

H1 has recently measured charm and beauty cross sections using a fit to the lifetime signature of charged particles in jets \cite{life}. This inclusive method yields measurements of differential cross sections that extend to larger values of transverse momenta than in previous HERA analyses, in which leptons from beauty quark decays were used to measure beauty cross sections. Fig. \ref{lifet} shows the measured cross sections as a function of the transverse momentum, $p_T^{jet1}$, of the leading jet. Taking into account the theoretical uncertainties, the beauty cross sections are consistent  both in normalisation and shape with a perturbative QCD calculation to next-to-leading order.
\noindent
In Fig. \ref{charm_dis}(right) a summary of the beauty cross section measurements from both H1 and ZEUS collaborations using different tagging methods is reported. At low transverse momentum values, the data tends to be slightly above the NLO QCD predictions. In this region the cross sections are extacted using double tagging techniques which allow to lower the kinematic threshold due to lower background. HERA II data will be needed to improve the cross section determination in the low- and high-$p_T$ region.

\section{$F_2^{c \bar{c}}$ and $F_2^{b \bar{b}}$}
\begin{figure}[h]
\begin{center}
\includegraphics[width=0.33\textwidth]{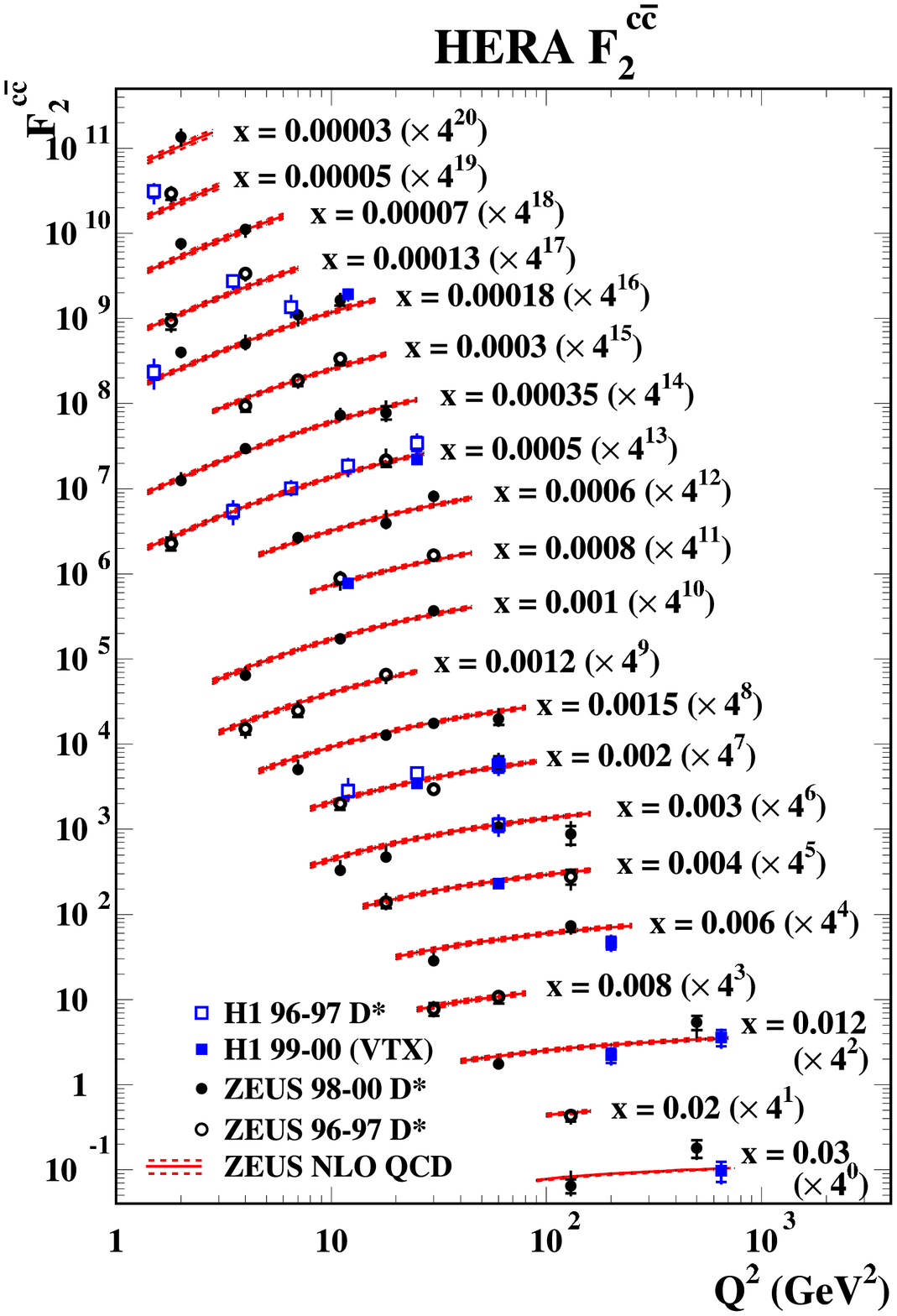}
\includegraphics[width=0.31\textwidth]{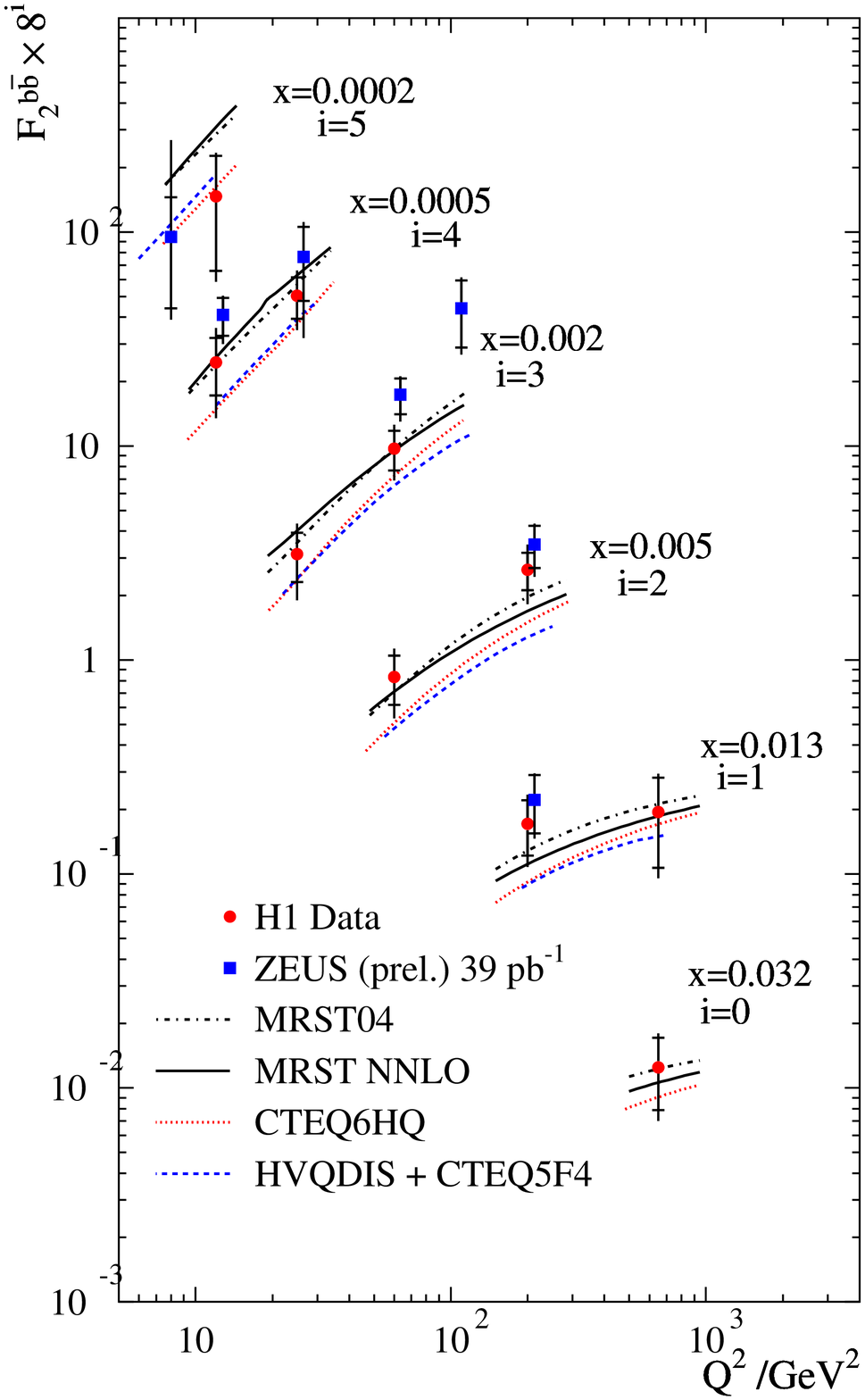}
\caption{(left) $F_2^{c \bar{c}}$ and (right) $F_2^{b \bar{b}}$ measurements as a function of $Q^2$ for different $x$ values.}
\label{struct}
\end{center}
\end{figure}

Measurements of the charm and beauty contributions to the inclusive structure function $F_2$ have been performed recently at HERA. The measurement of $F_2^{c \bar{c}}$ and $F_2^{b \bar{b}}$ has been done in a kinematic region where the extrapolation needed to correct for the full phase space is as small as possible. In Fig. \ref{struct}(left) a summary of the $F_2^{c \bar{c}}$ measurements as a function of $Q^2$ for different $x$ values is shown. The measurement of $F_2^{b \bar{b}}$ has been performed by ZEUS for the first time. In Fig. \ref{struct}(right) $F_2^{b \bar{b}}$ measured by the two experiments are compared with theoretical predictions, based on fixed-flavour and variable-flavour number schemes; a first NNLO calculation is also reported in the figure. The measurements from the two experiments are compatible within the errors and in agreement with the theory.

\section{Summary}
Recent results on beauty and charm production in {\it ep} collisions have been presented. 
The NLO QCD predictions describe well the charm data in a large range of $Q^2$, including the transition region between PHP and DIS. The beauty data agree with the NLO predictions at high-$p_T$ whilst at low-$p_T$ there is a tendency of the data to be slightly above the central NLO predictions.
The latest measurements of the $F_2^{c \bar{c}}$ and $F_2^{b \bar{b}}$ have been reported.

\end{document}